\documentclass[ amsmath,amssymb, showpacs, aps,twocolumn]{revtex4-1}

\usepackage{graphicx,bbm,amsmath}% Include figure files
\usepackage{dcolumn}% Align table columns on decimal point
\usepackage{bm}% bold math
\usepackage{color}

\begin{document}

\title{Backflash light characterization to prevent QKD zero-error hacking }

\author{A.~Meda$^a$}
\author{I.~P.~Degiovanni$^a$}
\author{A.~Tosi$^b$}
\author{Z.~L.~Yuan$^c$}
\author{G.~Brida$^a$}
\author{M.~Genovese$^a$}
\affiliation{$^a$INRIM, Strada delle Cacce 91, 10135 Torino, Italy}
\affiliation{$^b$Dipartimento di Elettronica, Informazione e Bioingegneria, Politecnico di Milano, Piazza Leonardo Da Vinci 32, 20133 Milano,
Italy}
\affiliation{$^c$ Toshiba Research Europe Ltd, 208 Cambridge Science Park, Cambridge CB4 0GZ, UK}

\begin{abstract}

Single photon avalanche diodes (SPADs) are the most commercially diffused solution for single-photon counting in quantum key distribution (QKD) applications. However, the secondary photon emission, arising from the avalanche of charge carriers during a photon detection, may be exploited by an eavesdropper to gain information without forcing errors in the transmission key.
In this paper, we characterise such backflash light in gated InGaAs/InP SPADs, and its spectral and temporal characterization for different detector models and different operating parameters. We qualitatively bound the maximum information leakage due to backflash light, and propose a solution.
 %using a single photon Optical Time Domain Reflectometer (OTDR) at telecom wavelength with low noise and high temporal resolution.
\end{abstract}

\maketitle
%\maketitle \section{ Introduction}

\emph{Introduction.} Quantum key distribution (QKD) is a method for sharing of secret cryptographic keys between two parties (Alice and Bob) with unprecedented level of security \cite{bb84, gis02, sca09}.
This level of security is assured by the laws of quantum mechanics and does not depend on technological resources available to an eavesdropper (Eve), provided that the QKD implementation does not deviate from its theoretical model. However, the security of practical systems (as any other cryptographic system) strongly depends on their device implementations. Deviation of QKD devices from their theoretical model can be exploited as side-channels or a back-doors \cite{xu10,lyd10}.

In 2010, two zero-error attacks on commercial QKD systems were reported exploiting defects in quantum signal encoding \cite{xu10} and detection \cite{lyd10}.
Shortly after, a plethora of quantum hacking attacks have been implemented with existing technologies to exploit devices imperfections in a number of QKD designs (different protocols, modules and systems) \cite{ger11, jai11,ger211, wei11, li11,jia12, wei12, mu12}. To guarantee security, each practical implementation must be carefully analyzed and tested against zero-error attacks.

 Single photon avalanche diodes (SPADs) are the most commercially-diffused solution for single photon detection in practical QKD implementation \cite{Pee09,Che10,Wan10,Stu11,Sas11,Yos13,Pat14,Kor15}. They can also be the most vulnerable components because they are optically exposed to Eve through the open quantum channel. Eavesdropper can either inject strong light to take control of these detectors to compromise the security of an entire QKD system. Alternatively, Eve could also measure passively any backflash light arising from avalanching carriers \cite{lac93} to learn the detected bit value (Fig. 1). Backflashes have been shown to exist in both InGaAs/InP and Si SPADs \cite{ace13b, kur01b}. However, these demonstrations are limited to free-space detectors and no experiments have been performed on fiber-pigtailed SPADs, which are the detectors of choice for all existing commercial QKD systems because of their practicality.

 Here, we present the first characterization of backflash light in fiber-pigtailed InGaAs SPADs from different manufacturers. We construct a reconfigurable optical time domain reflectometer (OTDR) operating in the single photon level \cite{hea80, war09, era10} with exceptional sensitivity. This OTDR allows an unambiguous identification of detector back-flashes from the conventional light back reflections, and provides a practical way to bound the information leakage, i.e., a fundamental step towards the QKD security. Furthermore, we showed that information can be leaked through back-flashes when two detector produce temporally distinguishable secondary emissions.

\begin{figure}[tbp]
\begin{center}
\includegraphics[angle=0, width=8 cm]{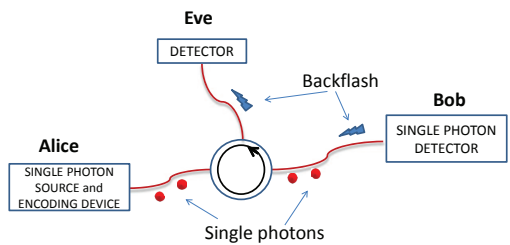}
\caption{Representation of an eavesdropper attack exploiting backflash light. Alice sends the photons of the key to Bob; when photons are revealed by Bob with a SPAD, a flash of light, the backflash, is emitted back to the channel. Eve can use a circulator to intercept this spot of light for acquiring information about the detector that has clicked.}
\label{bfs}
\end{center}
\end{figure}

\begin{figure}[tbp]
\begin{center}
\includegraphics[angle=0, width=8cm]{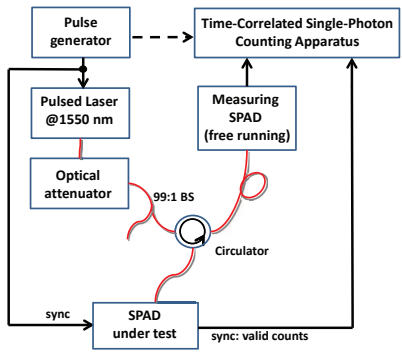}
\caption{A schematic representation of our experimental setup. A photon counting OTDR observes backflash light from SPADs under test. The source is an attenuated pulsed laser emitting at 1550 nm. Backflash light is detected by a free running InGaAs/InP detector. Time stamping of detected light is obtained by means of a correlator.}
\label{setupF}
\end{center}
\end{figure}

\emph{The experimental setup.}
The experimental setup used to analyze backflash light is depicted in Fig. \ref{setupF}. A strongly attenuated pulsed laser sends photons at 1550 nm to the InGaAs/InP SPAD under test (DUT). The back-reflected light is analysed with our photon-counting OTDR \cite{ras14, pia14} in order to quantify the amount of secondary emission photons which could serve as an information side-channel to Eve.

%Only By using photon counting technique the sensitivity can be greatly improved; the noise floor is established by the dark counts (and somehow also by the quantum efficiency) of %the photon counting detector, and the spatial resolution is set by the timing jitter of the source or detector.

The source is a commercial 1550 nm pulsed diode laser with pulse width of 300 ps and energy per pulse lower than 1 fJ. The laser output is sent to a single-mode optical fiber and attenuated to the single-photon level exploiting a fiber-coupled variable optical attenuator (max. attenuation 60 dB) combined with an additional 20 dB attenuation from a 99:1 fiber coupler.

We analyzed back-reflected and backflash light of two different InGaAs/InP detectors.
The first one, DUT1, is a prototype single photon detection module \cite{Tos12}; the second one, DUT2, is the commercial IdQuantique ID201, widely used in research laboratories. Both detectors are pigtailed and operate in gated mode.These devices are highly configurable in terms of detection efficiency, gate duration, and dead-time.  They also exploit active quenching and allow long avalanche durations (approximately 10 ns). Their configurability and long avalanche duration make them ideal to study backflashes.
The repetition rate of the laser pulses and of the DUTs was set with an external pulse generator to $f_{pg}$ = 50 kHz.
Both back-reflections and the DUTs' backflashes were directed by the circulator  to the measuring detector, a free running single-photon InGaAs/InP SPAD (IdQuantique ID220). The detector was operated in order to have low dark count rate (5 kHz), a nominal quantum efficiency of 10$\%$ and 130 ps timing resolution.
The output electrical signals from the OTDR detector as well as the one from the DUTs are sent to time-correlation-photon-counting (TCPC) electronics. Fig. 3 shows traces corresponding to the OTDR signals triggered by the laser pulses with an acquisition time of 60 minutes for DUT1 and DUT2 (a) and (b), respectively. The histogram represents the returned photons (due either to backflashes or back-reflections) as a function of time delay between the laser pulse emission and the detection by the OTDR detector. The horizontal axis depicts the time over which a detected photon has traveled.

\begin{figure}[tbp]
\begin{center}
\includegraphics[angle=0, width=8cm]{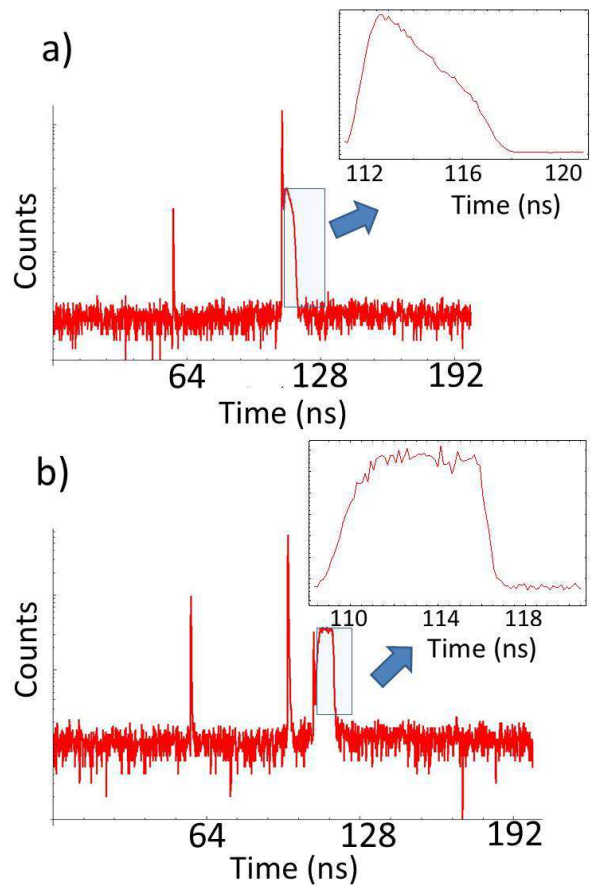}
\caption{Part a) and b) of the figure report the traces of the optical correlator after 60 minutes of acquisition for DUT1 and DUT2, respectively. It is visible the backflash peak, peculiar for each DUT, when an avalanche is triggered. We set, for DUT1, an excess bias voltage of 7 V, corresponding to a detection efficiency higher than 35 $\%$ and a gate width of 20 ns, while, for DUT2 the efficiency is $10\%$ and the gate width is 100 ns. Observing the zoom of the backflash peak for DUT1 and for DUT2, different shapes of the backflash peaks are evident. }
\label{bfF}
\end{center}
\end{figure}

In Fig. \ref{bfF} the sharp peaks arise from back-reflection at the connections between different different slices of fiber or between the fiber and other optical elements in the path (e.g. the circulator). There is also a rectangular and trapezoidal shaped feature that appears only when DUTs are switched on. We attribute this feature to backflash light emitted by the DUTs during avalanches.
Each type of DUTs has a unique, identifiable temporal profile, which gives away the information of the detector type and its manufacturer. We confirmed this finding by testing additional 4 devices of DUT1 type and 2  of DUT2 type.  Identifiable backflash profiles can be exploited by Eve to launch attacks tailored according to the detector type.

\emph{Photoemission characterization.} We evaluate the maximum possible information leakage $P_{L}$ due to backflash light for QKD systems implemented with detectors of type either DUT1 or DUT2. We consider a poorly designed QKD system which allows complete temporal discrimination of backflashes between different detectors. $P_{L}$ was estimated starting from the ratio between the number of detected backflashes, $N_{B}$ and the corresponding total number of valid counts  $N_{P}$  of the DUT. $N_{B}$ refers to only backflash events, i.e. after background subtraction. We considered the worst case scenario of the eavesdropper having an ideal equipment, i.e. lossless and with an ideal (unit) photon detection efficiency.  Thus $P_{L}$ was evaluated as
\begin{equation}\label{PL}
P_L =\frac{N_B}{N_P\cdot\eta_{det}\cdot\eta_{ch}}
\end{equation}
where corrections for losses and inefficiencies  of the OTDR system are applied, i.e. for the detection efficiency of the OTDR detector $\eta_{det}$, and for the losses in the optical channel connecting the DUT and the OTDR detector due to the circulator and the fiber connections, $\eta_{ch}$. To be conservative, we slightly overestimated losses and inefficiencies assuming $\eta_{ch} \eta_{det}$=0.05 on the basis of their approximate evaluations.
We obtain an information leakage $P_{L}$ of 9.8 $\%$ for DUT1 and of 6.0 $\%$ for DUT2. These results suggest that the information that Eve can get by observing backflash light is not negligible and countermeasures have to be in place.

\begin{figure}[tbp]
\begin{center}
\includegraphics[angle=0, width=8cm]{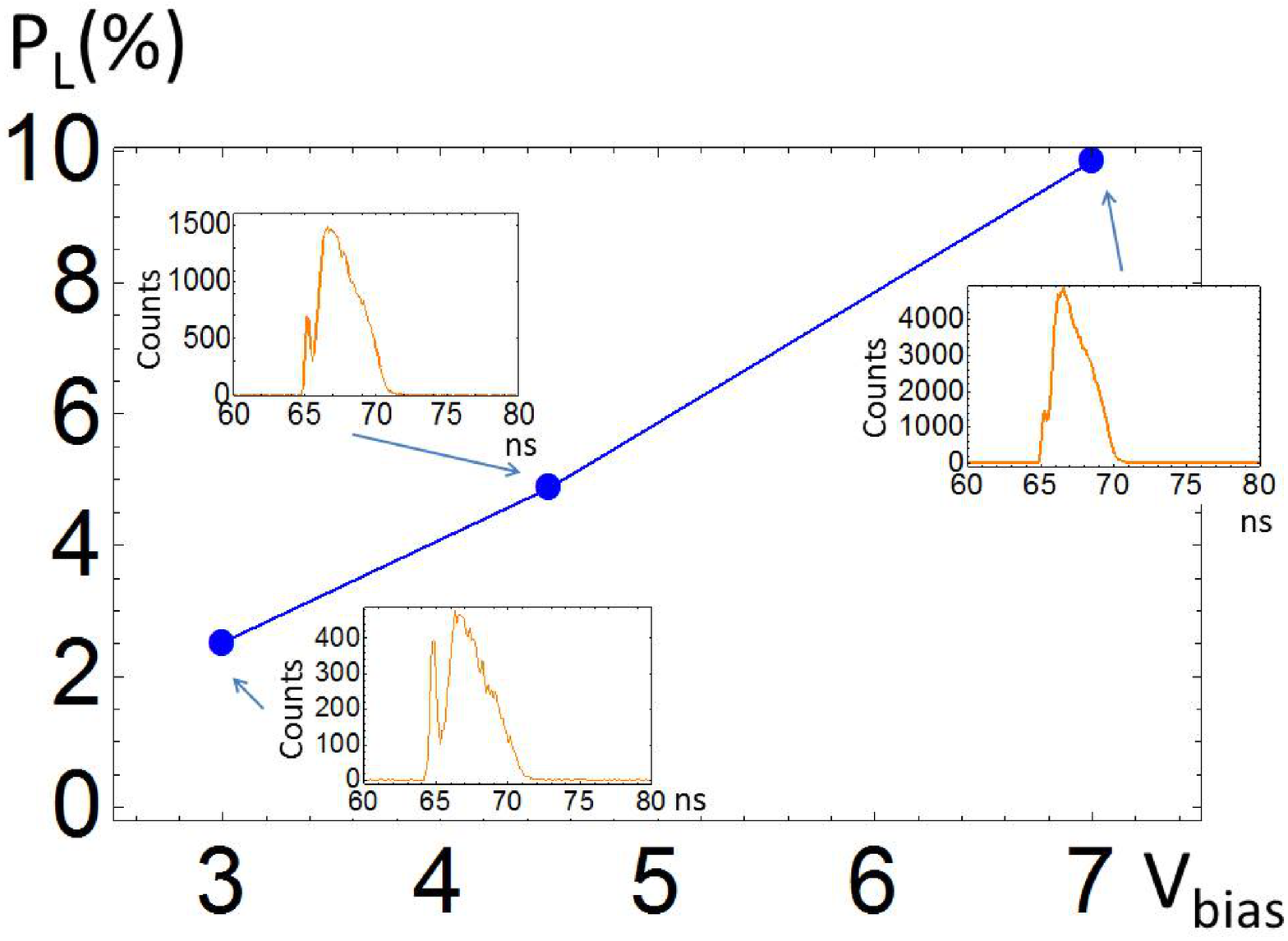}
\caption{Information leakage $P_L$ as a function of different excess bias voltages for the prototype detector DUT1. The peak of backreflected light is also reported: the smaller peak due to reflection of laser light on the diode surface is relatively more evident for low excess bias voltages. }
\label{bias}
\end{center}
\end{figure}

\begin{figure}[tbp]
\begin{center}
\includegraphics[angle=0, width=8cm]{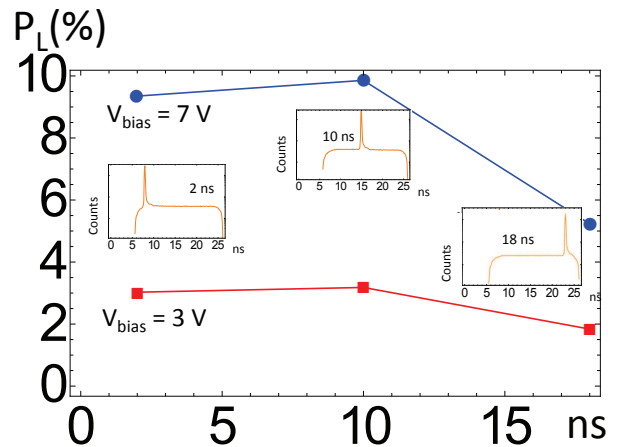}
\caption{Information leakage as a function of different arrival time of laser photons respect to the gate window that triggers the DUT.
The data refer to different bias voltage of DUT1, 7 V and 3V.}
\label{results}
\end{center}
\end{figure}

The backflash light is a consequence of carrier avalanches, which can be triggered by an absorbed photon when the device is biased beyond its breakdown voltage.  It is quenched, together with the avalanche itself  when the detector bias is lowered below breakdown. Thus, the intensity of the backflashes  strongly depends on the parameter setting of the quenching electronics. We investigated the information leakage percentage in DUT1 for different detector operating condition, i.e. varying detection efficiency, gate width etc. . The results are summarized in Fig. \ref{bias} and Fig. \ref{results}.
In Fig. \ref{bias}, information leakage as a function of different excess bias voltages for DUT1 is reported. We used three different settings of excess bias voltages, 3 V, 4.5 V and 7 V, corresponding to a nominal detection efficiency of 15\%, 22\% and 35\%, respectively. As shown, the backflash intensity increases with the excess bias of the detector, since the number of carriers also increases.

Fig. \ref{results} shows the information leakage as a function of the DUT gate delay relative to the incident laser pulses (2 $ns$, 10 $ns$ and 18 $ns$ after the beginning of the gating window). The two set of data refer to different bias voltage of DUT1: 7 V and 3V. A decrease of the information leakage is observed when the laser photons arrive at the end of the gating window. This is because late avalanches are quenched by the following edge of the gate window, rather than by the active quenching circuit.
The same effect explains the behavior observed when the laser peak is centered respect to the gating window but the latter presents different gating window width. Information leakage is reduced when a gating window comparable with the width of the temporal profile of backflash emission in DUT1 is used (e.g. 5 ns or less).

\begin{figure}[tbp]
\begin{center}
\includegraphics[angle=0, width=8cm]{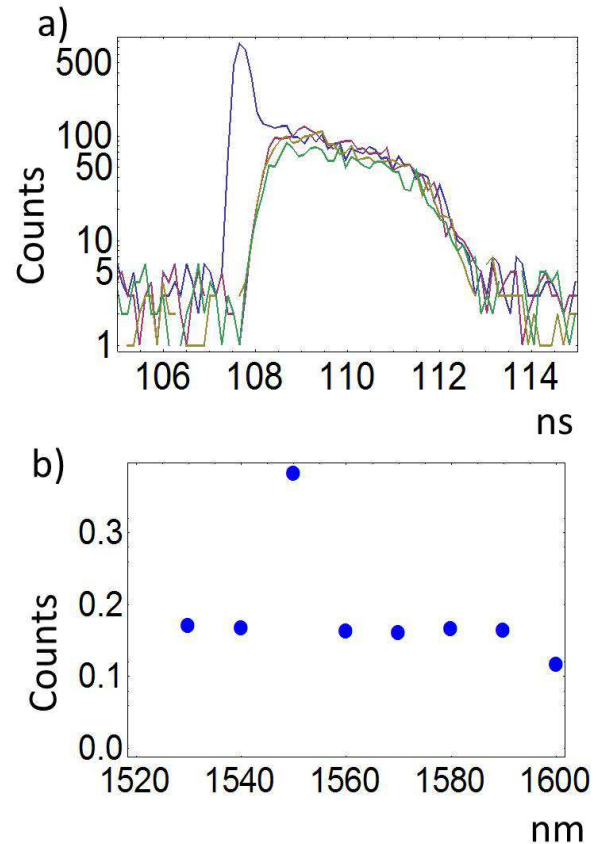}
\caption{Temporal distribution of the backflash counts after spectral filtering, respectively centered at 1530 nm (a)), 1550 nm (b)) and 1600 nm (c)). Part d) refers to the total counts of backflash light in the observed spectral range (from 1530 nm to 1600 nm). All measurements are performed with 10 nm bandwidth filters.}
\label{filter}
\end{center}
\end{figure}

To study the spectral distribution of the backflash emission, we integrated in our OTDR system a fiber optic tunable optical filter (Santec OTF-970) before the OTDR measuring detector. The filter spectral range is from 1530 nm to 1600 nm and we set 10 nm of pass bandwidth. Results are reported in Fig. \ref{filter}a: the four reported profiles are the temporal distribution of the backflash counts, respectively centered at 1530 nm, 1550 nm, 1570 nm  and 1600 nm. The temporal emission profile is similar to the one obtained without spectral filtering (Fig. \ref{bfF}a) for all the wavelengths. When the filter is centered on 1550 nm, the reflection peak dominates.

Fig. \ref{filter} (b) reports the total backflash counts as a function of the filter center wavelength. The subtraction of backreflected light is performed measuring the laser light backreflected by DUT1 with bias voltage applied but in absence of gate signal.
Backflash emission is broadband, at least beyond the spectral range of our tunable filter, because it is due to the relaxation of hot carriers generated in the multiplication region \cite{Lac93, Ace13}. In the spectral region of our tunable filter it is reasonably uniform, except for the region around the 1550 nm (1545 - 1555 nm) where a peak is observed even after the subtraction of the laser light backreflected by the DUT (see the sharp peak in Fig. \ref{filter}a). It is reasonably expected that the sharp peak present even after the background subtraction is due to backreflected laser light, since we observed that the reflectivity of the diode varies with the applied bias \cite{footnote} and we attribute this to the refractive index change in the semiconductor material \cite{Ben90}.

This is confirmed by the measurement of backflashes spectrum performed with a pulsed laser at 1570 nm as source of our spectrally filtered OTDR. In this configuration we expected to observe the sharp peak disappearing at 1550 nm, and appearing at 1570 nm, and, indeed this was exactly what happened, confirming that the sharp peak was just due to the change of reflectivity of the SPAD surface due to its bias voltage change.

\emph{Conclusion.}
We proved that backflashes are significant in commercial InGaAs/InP single photon detector operating at telecom wavelength. These backflashes could potentially allow  a serious security breach in a poorly designed QKD system. Proper design and testing of QKD systems should be implemented in order to avoid attacks based on backflashes. Possible solutions can be based on passive optical devices \cite{Luc15} such as isolators, circulators or spectral filters to prevent backflashes leaking out of the QKD system. We underline that these countermeasures should consider the wide bandwidth of the backflash light emission. Careful characterisation of the spectral behavior of this optical components are necessary to ensure their operation as countermeasures.

Following this line of thought, a combination of circulators or isolators with interference optical filters at the input of the QKD system should essentially nullify the information leakage due to backflash light at the cost of some additional optical loss (the insertion losses of the optical filter and of the circulator) on the QKD signal. Furthermore, as discussed in connection with Fig. \ref{results}, the use of as short as possible gates and small avalanche will reduce the emitted backflash light. In this sense, fast-gated detectors \cite{Nam06, You07, Dix09, Lia12, Res13, Sca15} represent an interesting solution for QKD systems not only in terms of speed but also for their much lower avalanche charges (even 100 times lower!). In fact, it is expected that they produce a significant low backflash light emission. In addition, use of short gates make it harder for Eve for temporal discrimination of backflash light. Testing backflashes of fast gated detectors is an interesting research direction.

For QKD application, superconducting nanowire single-photon detector are an excellent option. Indeed, in addition to their high detection efficiency, their low dark-count rate, and  their short recovery time \cite{Mar15}, is is expected that they do not present any backflash light (i.e. any related information leakage). Unfortunately, they need cryogenic temperature of operation and because of the high cost of cryogenic equipment they appear at the moment not suitable for practical deployment of QKD systems in the real world.

%\section*{Aknowledgments}

\textbf{Aknowledgments.}
This work has received funding from the European Union's Horizon 2020 and the EMPIR and EMRP Participating States in the context of the project EXL02 SIQUTE and 14IND05 MIQC2 respctively. We acknowledge funding support also from FIRB Project No. D11J11000450001 funded by MIUR, and by the NATO SPS 984397.


\begin{thebibliography}{99}

\bibitem{bb84} C. H. Bennett and G. Brassard, Conference on Comput-
ers, Systems and Signal Processing, Bangalore (India),
175 (1984).

\bibitem{gis02} N. Gisin, G. Ribordy, W. Tittel, and H. Zbindenr, Rev.
of Mod. Phys. \textbf{74}, 145 (2002).

\bibitem{sca09} V. Scarani, H. Bechmann-Pasquinucci, N. J. Cerf,
M. Dusek, N. L\"ukenhaus, and M. Peev, Rev. of Mod.
Phys. \textbf{81}, 1301 (2009).

%\bibitem{KLo14} H-K Lo,	M. Curty, and K. Tamaki, Nature Phot. \textbf{8}, 595 (2014).


\bibitem{xu10} C. C. W. Lim, M. Curty, N. Walenta, F. Xu, and
H. Zbinden, Phys. Rev. A \textbf{89}, 022307 (2014).

\bibitem{lyd10} L. Lydersen, C. Wiechers, C. Wittmann, D. Elser, J. Skaar, and V. Makarov, Nature Photonics
\textbf{4}, 686 (2010).

\bibitem{ger11} I. Gerhardt, Q. Liu, A. Lamas-Linares, J. Skaar, V. Scarani, V. Makarov, and C. Kurtsiefer,
Phys. Rev. Lett. \textbf{107}, 170404 (2011)


\bibitem{jai11} N. Jain, C. Wittmann, L. Lydersen, C. Wiechers, D. Elser, C. Marquardt, V. Makarov, and
G. Leuchs, Phys. Rev. Lett. \textbf{107}, 110501 (2011).

\bibitem{ger211} I. Gerhardt, Q. Liu, A. Lamas-Linares, J. Skaar, C. Kurtsiefer, and V. Makarov, Nature
Communications \textbf{2}, 349 (2011).

\bibitem{wei11} H. Weier, H. Krauss, M. Rau, M. Frst, S. Nauerth, and H. Weinfurter, New J. Phys. \textbf{13},
073024 (2011).

\bibitem{li11} H.-W. Li, S. Wang, J.-Z. Huang, W. Chen, Z.-Q. Yin, F.-Y. Li, Z. Zhou, D. Liu, Y. Zhang,
G.-C. Guo, W.-S. Bao, and Z.-F. Han, Phys. Rev. A \textbf{84}, 062308 (2011).

\bibitem{jia12} M.-S. Jiang, S.-H. Sun, C.-Y. Li, and L.-M. Liang, Phys. Rev. A \textbf{86}, 032310 (2012).

\bibitem{wei12} H.-W. Li, S. Wang, J.-Z. Huang, W. Chen, Z.-Q. Yin, F.-Y. Li, Z. Zhou, D. Liu, Y. Zhang,
G.-C. Guo, W.-S. Bao, and Z.-F. Han, Phys. Rev. A \textbf{84}, 062308 (2011).

\bibitem{mu12} M.-S. Jiang, S.-H. Sun, C.-Y. Li, and L.-M. Liang, Phys. Rev. A \textbf{86}, 032310 (2012).

%\bibitem{had09} R. Hadfield, Nature Photonics \textbf{3}, 696 (2009).

%\bibitem{mig11}A. L. Migdall, I. P. Degiovanni, J. Y. Cheung, S. V. Polyakov, and J. Fanv, Journ. of Mod.
%Opt. \textbf{58}, 169 (2011).

%\bibitem{eke91} A. K. Ekert, Phys. Rev. Lett. 67, 661 (1991).

%\bibitem{ben92} C. Bennett, Phys. Rev. Lett. 68, 3121 (1992).

%\bibitem{aci04} A. Acin, N. Gisin, and V. Scarani, Phys. Rev. A 69, 012309 (2004).

%\bibitem{ino02} K. Inoue, E. Waks, and Y. Yamamoto, Phys. Rev. Lett. 89, 037902 (2002).

%\bibitem{tak05} H. Takesue, E. Diamanti, T. Honjo, C. Langrock, M. M. Fejer, K. Inoue, and Y. Yamamoto,
%New J. of Phys. 7, 232 (2005).

%\bibitem{bra07} C. Branciard, N. Gisin, N. Lutkenhaus, and V. Scarani, Quantum Information Computation
%7, 639 (2007).

%\bibitem{stu09} D. Stucki, C. Barreiro, S. Fasel, J. Gautier, O. Gay, N. Gisin, R. Thew, Y. Thoma, P. Trinkler,
%xF. Vannel, and H. Zbinden, Opt. Exp. 17, 13326 (2009).


\bibitem{Pee09} M. Peev, C. Pacher, R. All\'eaume, C. Barreiro, J. Bouda, W. Boxleitner,T. Debuisschert, E. Diamanti, M. Dianati, J. F.
Dynes, S. Fasel, S. Fossier, M. Fürst, J-D. Gautier, O. Gay, N. Gisin, P. Grangier, A. Happe, Y. Hasani, M. Hentschel, H. H\"ubel, G. Humer, T. L\"anger, M. Legr\'e, R. Lieger, J. Lodewyck, T. Lor\"unser, N. L\"utkenhaus, A. Marhold, T. Matyus, O. Maurhart, L. Monat, S. Nauerth, J-B. Page, A. Poppe, E. Querasser, G. Ribordy, S. Robyr, L. Salvail, A. W. Sharpe, A. J.
Shields, D. Stucki, M. Suda, C. Tamas, T. Themel, R. T. Thew, Y. Thoma, A. Treiber, P. Trinkler, R. Tualle-Brouri, F. Vannel, N. Walenta, H. Weier, H. Weinfurter, I. Wimberger, Z. L. Yuan, H. Zbinden, A. Zeilinger, New J. Phys. \textbf{11}, 075001 (2009).

\bibitem{Che10} T-Y. Chen, J. Wang, H. Liang, W-Y. Liu, Y. Liu, X. Jiang, Y. Wang, X. Wan, W-Q. Cai, L. Ju, L-K Chen, L-J. Wang, Y. Gao, K. Chen, C-Z. Peng, Z-B Chen, and J-W. Pan, Opt. Express \textbf{18}, 27217 (2010).


\bibitem{Wan10}S. Wang, W. Chen, Z-Q. Yin, Y. Zhang, T. Zhang, H-W. Li, F-X. Xu, Z. Zhou, Y. Yang, D-J. Huang, L-J. Zhang, F-Y. Li, D. Liu, Y-G. Wang, G-C. Guo, and Z-F. Han, Opt. Lett. \textbf{35}, 2454 (2010).


\bibitem{Stu11} D. Stucki, M. Legr\'e, F. Buntschu, B. Clausen, N. Felber, N. Gisin, L. Henzen, P. Junod, G. Litzistorf, P. Monbaron, L. Monat, J-B. Page, D. Perroud, G. Ribordy, A. Rochas, S. Robyr, J. Tavares, R. Thew, P. Trinkler, S. Ventura,  R. Voirol, N. Walenta, and H. Zbinden, New J. Phys. \textbf{13}, 123001 (2011).

\bibitem{Sas11} M. Sasaki, M. Fujiwara, H. Ishizuka, W. Klaus, K. Wakui, M. Takeoka, S. Miki, T. Yamashita, Z. Wang, A. Tanaka, K. Yoshino, Y. Nambu, S. Takahashi, A. Tajima, A. Tomita, T. Domeki, T. Hasegawa, Y. Sakai, H. Kobayashi, T. Asai, K. Shimizu, T. Tokura, T. Tsurumaru, M. Matsui, T. Honjo, K. Tamaki, H. Takesue, Y. Tokura, J. F. Dynes, A. R. Dixon, A. W. Sharpe, Z. L. Yuan, A. J. Shields, S. Uchikoga, M. Legr\'e, S. Robyr, P. Trinkler, L. Monat, J.-B. Page, G. Ribordy, A. Poppe, A. Allacher, O. Maurhart, T. L\"anger, M. Peev, and A. Zeilinger, Opt. Express \textbf{19}, 10387 (2011).

%\bibitem{Fro13} B. Fröhlich, et al., Nature \textbf{501}, 69 (2013).

\bibitem{Yos13} K. Yoshino, T. Ochi, M. Fujiwara, M. Sasaki and A. Tajima, Opt. Express \textbf{21}, 31395 (2013).

\bibitem{Pat14} K. A. Patel, J. F. Dynes, M. Lucamarini, I. Choi, A. W. Sharpe, Z. L. Yuan, R. V. Penty, and A. J. Shields, Appl. Phys. Lett. \textbf{104}, 051123 (2014).

\bibitem{Kor15} B. Korzh, C. C. W. Lim,	R. Houlmann,	N. Gisin,	M. J. Li,	D. Nolan,	B. Sanguinetti,	R. Thew, and H. Zbinden, Nature Phot. \textbf{9}, 163 (2015).






\bibitem{lac93} A. Lacaita, F. Zappa, S. Bigliardi, and M. Manfredi, IEEE Transactions
on Electron Devices,  \textbf{40}, 577 (1993).

\bibitem{ace13b} F. Acerbi, A. Tosi, and F. Zappa, IEEE Photonics Technol. Lett. \textbf{25}, 18 (2013).


\bibitem{kur01b} C. Kurtsiefer, P. Zarda, S. Mayer, and H. Weinfurter, Journ. Mod. Opt. \textbf{48}, 13 (2001).





\bibitem{hea80} P. Healey and P. Hensel, Electr. Lett. \textbf{16}, 16 (1980).

\bibitem{war09} R. Warburton, M. Itzler, and G. Buller, Appl. Phys. Lett. \textbf{94}, 071116 (2009).

\bibitem{era10} P. Eraerds, M. Legr\'e, J. Zhang, H. Zbinden, and N. Gisin, J. Light. Tech. \textbf{28}, 952 (2010).

\bibitem{ras14} M. L. Rastello, I. P. Degiovanni, A. G. Sinclair, S. Kuck, C. J. Chunnilall, G. Porrovecchio,
M. Smid, F. Manoocheri, E. Ikonen, T. Kubarsepp, D. Stucki, K. S. Hong, S. K. Kim, A. Tosi,
G. Brida, A. Meda, F. Piacentini, P. Traina, A. A. Natsheh, J. Y. Cheung, I. M\"uller, R. Klein,
and A. Vaigu, Metrologia \textbf{51}, S267 (2014).

\bibitem{pia14} F. Piacentini, A. Meda, P. Traina, Hong Kee Suk, I. P. Degiovanni, G. Brida, M. Gramegna, I. Ruo Berchera, M. Genovese, M. L. Rastello, Int. J. Quantum Inform. \textbf{12}, 1461014 (2014).

\bibitem{Tos12} A. Tosi, A. Della Frera, A. Bahgat Shehata, and C. Scarcella, Rev. Sci. Instrum., \textbf{83}, 013104 (2012).

\bibitem{Lac93}  A. Lacaita, F. Zappa, S. Bigliardi, and M. Manfredi, IEEE Trans. Electron Devices, \textbf{40},  577 (1993).

\bibitem{Ace13} F. Acerbi, A. Tosi, and F. Zappa, IEEE Photonics Technol. Lett. \textbf{25}, 1778 (2013).

\bibitem{footnote} In particular, a relative in reflectivity increase of almost one order of magnitude of the SPAD surface was observed in the case of non-polarised versus polarised (but non-gated) detector.

\bibitem{Ben90} B. R. Bennett, R. A. Soref and J. A. Del Alamo, IEEE J. of Quantum Electronics, \textbf{26} 113 (1990).


\bibitem{Luc15} M. Lucamarini, I. Choi, M. B. Ward, J. F. Dynes, Z. L. Yuan, adn A. J. Shields, Phys. Rev. X 5, 031030 (2015).



%\bibitem{leg07} M. Legre, R. Thew, H. Zbinden, and N. Gisin, Opt. Express \textbf{15}, 8237 (2007).

\bibitem{Nam06}  N. Namekata, S. Sasamori, and S. Inoue, Opt. Express 14, 10043 (2006).

\bibitem{You07} Z. L. Yuan, B. E. Kardynal, A. W. Sharpe and A. J. Shields, App. Phys. Lett. \textbf{91}, 041114 (2007).


\bibitem{Dix09} A. R. Dixon, J. F. Dynes, Z. L. Yuan, A. W. Sharpe, A. J. Bennett, and A. J. Shields, Appl. Phys. Lett. \textbf{94}, 231113 (2009).

%\bibitem{Zha09} J. Zhang, R. Thew, C. Barreiro, H. Zbinden, Appl. Phys. Lett. \textbf{95}, 091103 (2009).

\bibitem{Lia12} X.-L. Liang, J.-H. Liu, Q. Wang, D-B. Du, J. Ma, G. Jin, Z.-B. Chen, J. Zhang, and J.-W. Pan, Rev. Sci. Instrum. \textbf{83}, 083111 (2012).

\bibitem{Res13} A. Restelli, J. C. Bienfang, and A. L. Migdall,  Appl. Phys. Lett. \textbf{102}, 141104 (2013)

\bibitem {Sca15} C. Scarcella, G. Boso, A. Ruggeri, and A. Tosi, IEEE J. Sel. Top. Quantum Electron. \textbf{21}, 1 (2015).


\bibitem{Mar15} F. Marsili,	V. B. Verma, J. A. Stern,	S. Harrington,	A. E. Lita,	T. Gerrits,	I. Vayshenker,	B. Baek,	M. D. Shaw,	R. P. Mirin, and S. W. Nam, Nature Photon. \textbf{7}, 210 (2013).









\end{thebibliography}
\end{document}